\documentstyle[12pt,epsf]{article}
\begin{document}
%\pagenumbering{empty}

% The following definitions need to be customised;

% Will appear on page headings
%\authorrunning{G. Boyd, B. All\'es, M. D'Elia, A. Di Giacomo}
%\titlerunning{{\talknumber}: Topology in QCD}
 
% Now the full name of author and talk

% For plenary talks, the talk number is that of the session
\def\talknumber{1503} 

\centerline{{\talknumber}: {\bf Topology in QCD}}

\vskip 1cm

G. Boyd$^{\rm a}$, (boyd@rccp.tsukuba.ac.jp) 

B. All\'es$^{\rm b}$, (alles@sunmite.mi.infn.it) 

M. D'Elia$^{\rm c\, d}$, (delia@dirac.ns.ucy.ac.cy) 

A. Di Giacomo$^{\rm d}$, (digiaco@mailbox.difi.unipi.it)

%\maketitle

\vskip 5mm

$^{\rm a}$Centre for Computational Physics, 
University of Tsukuba, Japan 

$^{\rm b}$Dipartimento di Fisica, Sezione Teorica, 
Universit\`a di Milano, Italy 

$^{\rm c}$Department of Natural Sciences, 
University of Cyprus, Cyprus

$^{\rm d}$Dipartimento di Fisica, Universit\`a di Pisa, Italy

\begin{abstract}
Topology on the lattice is reviewed. In quenched QCD topological
susceptibility $\chi$ is fully understood. The Witten-Veneziano mechanism
for the $\eta'$ mass is confirmed. The topological susceptibility
drops to zero at the deconfining phase transition. Preliminary results
are also presented for $\chi$ and $\chi'$ in full QCD, and for the
spin content of the proton. The only problem there is the difficulty of
the usual Hybrid Monte Carlo algorithm to bring topology to equilibrium.
\end{abstract}
\section{Introduction}
Topology plays a fundamental role in QCD. The key equation is the
$U_A(1)$ anomaly
\begin{equation}
 \partial_\mu J_\mu^5 = - 2 N_f Q(x)
\label{eq:anomaly}
\end{equation}
where $J^5_\mu=\sum_{i=1}^{N_f} \overline{\psi}_i \gamma^5\gamma_\mu
\psi_i$ is the singlet axial current, $N_f$ the number of light 
flavours and
\begin{equation}
 Q(x) = {g^2 \over {64 \pi^2}} F_{\mu\nu}^a F_{\rho\sigma}^a 
 \epsilon_{\mu\nu\rho\sigma}
\label{eq:Q}
\end{equation}
is the topological charge density. $Q(x)$ is related to the Chern
current $K_\mu(x)$
\begin{equation}
 \partial_\mu K_\mu(x) = Q(x)
\label{eq:k}
\end{equation}
with
\begin{equation}
 K_\mu = {g^2 \over {16 \pi^2}} \epsilon_{\mu\nu\rho\sigma}
 A^a_\nu \left(\partial_\rho A^a_\sigma - {1 \over 3}
 g f^{abc} A^b_\rho A^c_\sigma \right)
\label{eq:kmu}
\end{equation}
and as a consequence $Q=\int Q(x)$ is an integer on smooth classical
configurations with finite action. Eqs. (\ref{eq:anomaly}-\ref{eq:k}) yield
\begin{equation}
 \partial_\mu \left( J^5_\mu(x) + 2 N_f K_\mu(x) \right) = 0 
\label{eq:noanomaly}
\end{equation}
whence Ward identities can be derived.
At the leading order in the $1/N_c$ expansion, ($N_c \longrightarrow \infty$,
with $g^2 N_f$ fixed), $Q(x)$ is zero, $U_A(1)$ is a symmetry and 
$\eta'$ is its Goldstone particle, $m_{\eta'} =0 $. The anomaly acts
as a perturbation and shifts the position of the pole from zero to
the actual $\eta'$ mass. From Eq.(\ref{eq:noanomaly}) by inserting the
quark mass terms, the relation follows~\cite{witten,veneziano}
\begin{equation}
 \chi = {{f_\pi^2} \over {2 N_f}} 
 \left( m_\eta^2 + m_{\eta'}^2 - 2 m_K^2 \right)
\label{eq:mass}
\end{equation}
where $\chi$ is the topological susceptibility of the unperturbed 
vacuum $(1/N_c = 0)$ defined as
\begin{equation}
 \chi = \int \hbox{d}^4(x-y) \partial^x_\mu \partial^y_\nu 
 \langle 0 | T(K_\mu(x) K_\nu(y)) | 0 \rangle _{\rm quenched}.
\label{eq:chi}
\end{equation}
The subscript ``quenched'' indicates that the matrix element
has to be computed on the ground state of the $1/N_c = 0$ theory.
In particular this implies that fermion loops, which are
$O(g^2 N_f)$ are put to zero. Eq.(\ref{eq:mass}) gives
\begin{equation}
 \chi = (180\;\hbox{MeV})^4 
\label{eq:180}
\end{equation}
which is expected to be valid within an order $O(1/N_c)$ of accuracy.
Eq.(\ref{eq:mass}) is a peculiar equation relating physical quantities
(masses, $f_\pi$, $N_f$) to $\chi$, which exists in an artificial 
$1/N_c=0$ world. Its verification, however, is a check of the validity
of the expansion, which is a fundamental issue. Lattice is an ideal
tool to produce this artificial world, and in particular the absence
of fermions in it simplifies the numerical work. Eq.(\ref{eq:chi})
uniquely fixes the prescription for the singularity in the product
of operators $K_\mu(x) K_\nu(y)$ as $x \longrightarrow y$: $\delta$-like
singularities disappear after integration and this uniquely determines
$\chi$. Eq.(\ref{eq:chi}) is a specific prescription for the notation
\begin{equation}
 \chi=\int \hbox{d}^4 x \langle 0 | T(Q(x) Q(0) ) | 0 \rangle_{\rm quenched}.
\label{eq:usualchi}
\end{equation}
When computing $\chi$ by any regularization scheme, like lattice,
an appropriate subtraction must be performed to satisfy the prescription
Eq.(\ref{eq:chi}). 

The behaviour of $\chi$ at finite temperature, and more specifically,
at deconfinement, is an important key to understand the structure of
QCD vacuum~\cite{shuryak}

%\section{Determining $\chi$ from lattice}
A regularized version of the operator $Q(x)$, $Q_L(x)$, can be defined
on the lattice. There is a large arbitrariness in this definition, by
terms of higher order in the lattice-spacing which go to zero in the
continuum limit. In general $Q_L(x)$ will not be an exact divergence,
so that a multiplicative renormalization $Z$ with respect to continuum
can exist~\cite{campo}. If $a$ is the lattice-spacing,
\begin{equation}
 Q_L(x) = a^4 Z Q(x) + O(a^6).
\label{eq:limit}
\end{equation}
A lattice susceptibility $\chi_L$ is defined as 
\begin{equation}
 \chi_L = \sum_x \langle Q_L(x) Q_L(0) \rangle = {{\langle Q_L^2 
 \rangle}\over V}.
\label{eq:chilattice}
\end{equation}
In general the definition (\ref{eq:chilattice}) will not satisfy
the prescription Eq. (\ref{eq:chi}). It will be~\cite{vicari1}
\begin{equation}
 \chi_L = a^4 Z^2 \chi + M + O(a^6) 
\label{eq:renor}
\end{equation}
where $M$ is a mixing with the continuum operators having
dimension $\leq 4$ ($\overline{\beta}$ is the beta function)
\begin{equation}
 M(\beta) = A(\beta) \langle {{\overline\beta(g)} \over g}
                      F^a_{\mu\nu} F^a_{\mu\nu} \rangle a^4 +
            P(\beta) \cdot 1 .
\label{eq:mixing}
\end{equation}
 From Eq.(\ref{eq:renor}) 
\begin{equation}
 \chi = {{\chi_L - M} \over {Z^2 a^4}}.
\label{eq:lattchi}
\end{equation}
$\chi_L$ is measured on the lattice numerically. $a^4$ is determined 
as usual by comparison to a physical quantity ($\rho$ mass, string
tension); $M$ and $Z$ can be determined non-perturbatively by a 
procedure known as heating~\cite{vicari2}. 
The idea is that classical configurations
with known topological charge, can be dressed by quantum fluctuations
without modifying the topological content since topological charge 
is difficult to change by the usual local Monte Carlo algorithms.
$Z$ is determined by measuring the total charge $Q_L$ on a configuration with
an instanton, where $Q=1$, Eq.(\ref{eq:limit}). $M$ is determined
from Eq.(\ref{eq:renor}) by measuring $\chi_L$ 
on the sector $Q=0$, where as a consequence of
Eq.(\ref{eq:chi}), $\chi=0$.

In Eq.(\ref{eq:lattchi}) $\chi_L$, $M$ and $Z$ strongly depend on the
choice of $Q_L$, on the action and on the coupling constant 
$\beta=2 N_c / g^2$. $a$ depends on the action and on $\beta$.
$\chi$ must be independent of all these parameters. Fig. 1 shows
$\chi$ for $SU(3)$~\cite{a1}, determined for several $\beta$'s and by use of
different operators: as visible in the figure
%\begin{equation}
 $\chi = (175\pm 5$ MeV)$^4$.
%\label{eq:chi3}
%\end{equation}
 For $SU(2)$ (Fig. 2)~\cite{a2} $\chi$ is somewhat larger:
%\begin{equation}
 $\chi = (198\pm 6$ MeV)$^4$.
%\label{eq:chi2}
%\end{equation}

\begin{figure}
\vspace{2.7cm}
\includegraphics{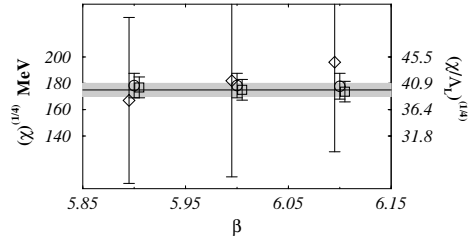} 
\caption{$\chi$ for $SU(3)$. Diamonds, circles and squares
correspond to the 0, 1 and 2-smeared operators.}
\end{figure}

\begin{figure}
\vspace{2.7cm}
\includegraphics{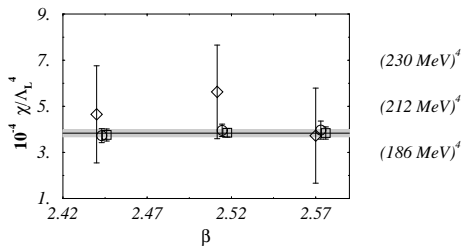} 
\caption{$\chi$ for $SU(2)$. Diamonds, circles and squares
correspond to the 0, 1 and 2-smeared operators.}
\end{figure}

The determination by using the so-called geometrical method, if
accompained by the appropriate subtraction, agrees with the other
choices of $Q_L$.

 Fig. 3 shows the behaviour of $\chi$ across deconfinement. The drop
is stronger for $SU(3)$ than for $SU(2)$~\cite{a2}.

\begin{figure}
\vspace{2.7cm}
\includegraphics{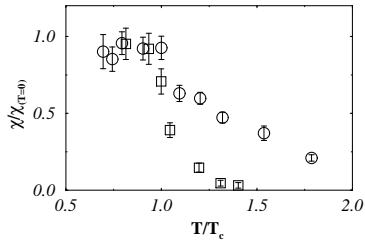} 
\caption{$\chi$ across the deconfinement transition for 
$SU(2)$ (circles) and $SU(3)$ (squares).}
\end{figure}

\section{Full QCD}
By use of the same procedure described in the previous section,
$\chi$ can be determined in full QCD. The expectation from the Ward
identities is that 
\begin{equation}
 \chi \approx \langle \sum_{i=1}^{N_f} m_i \overline{\psi}_i \psi_i
 \rangle + O(m_q^2).
\label{eq:ward}
\end{equation}
Our preliminary result, extracted from simulating at $\beta=5.35$
where $a=0.11(1)$ fm with 4 staggered fermions at $am=0.01$ is
\begin{equation}
 \chi = (110 \pm 8 \; \hbox{MeV})^4
\label{eq:chifermion}
\end{equation}
to be compared to the predicted value 
\begin{equation}
 \chi = {m_q \over N_f} \langle \overline{\psi} \psi\rangle _{m=0}
 \sim (109 \; \hbox{MeV})^4 .
\end{equation}
On the same sample of configurations we obtain for $\chi'$ the 
preliminary value
\begin{equation}
 \chi' = 258 \pm 100 \; \hbox{MeV}^2 \;\;\;\;
 \hbox{or} \;\;\; \sqrt{\chi'} = 19 \pm 4 \; \hbox{MeV}
\label{eq:258}
\end{equation}
which is compatible with the value expected from sum rules~\cite{narison}
$\sqrt{\chi'} = 25 \pm 3 \; \hbox{MeV}$.
However both these determinations are preliminary because of
the effect shown in Fig. 4 where we display the history of the 
topological charge $Q$ along the Monte Carlo updating which produces
the configurations~\cite{a3}. 
In the updating algorithms used in quenched QCD
(Metropolis, heat-bath) the topological charge has tipically $50-100$
steps of authocorrelation time. It thermalizes slowly with respect to local
quantum fluctuations (and this is the basic property which allows the
heating method for the measurement of $Z$ and $M$ as explained in
section 1), but a thermalized sample of configurations can be 
prepared in a reasonable CPU time. The algorithm used with dynamical
fermions, the hybrid Monte Carlo, performs very badly in that respect,
as is visible from Fig. 4. The configurations there correspond to
about 700 CPU hours of an APE Quadrics with 25 GFlop which is a huge
time. Our sample has thus a much smaller number of independent
configurations than shown in Fig. 4, and therefore the errors
in the results given in Eqs. (\ref{eq:chifermion}) 
and (\ref{eq:258}) are underestimated.

\begin{figure}
\vspace{2.7cm}
\includegraphics{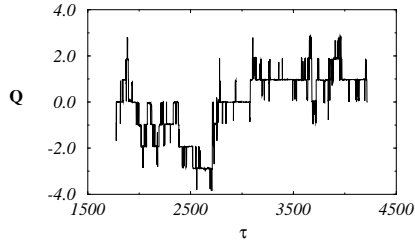} 
\caption{History of $Q$ in a Hybrid Monte Carlo run as a
function of the molecular dynamic time $\tau$.}
\end{figure}

The same uncertainty affects our determination, on the same sample,
of the spin content of the proton.
%\section{Spin content of the proton}
The matrix element of $J^5_\mu$ between proton states can be
parametrized as 
\begin{equation}
 \langle {\vec p'}, s' | J^5_\mu | {\vec p},s \rangle =
 \overline{u}({\vec p'}, s') \left( G_1(k^2) \gamma_\mu
 \gamma^5 - G_2(k^2) \gamma^5 k_\mu \right) u({\vec p}, s)
\label{eq:matrix}
\end{equation}
with $k=p - p'$. The form factor $G_1$ is related to the so-called
spin content of the proton $\Delta\Sigma$, $G_1(0) = \Delta\Sigma$
where $\Delta\Sigma \equiv \Delta u + \Delta d +\Delta s$
is the contribution of the different quarks species to the spin of the
proton. The na\"{\i}ve expectation would be $ \Delta\Sigma \sim 0.7$.
The value determined from the moments of the spin dependent structure
functions of inelastic scattering of leptons on nucleons is much
lower: $\Delta\Sigma =0.2(1)$. The lattice allows a determination
of $\Delta\Sigma$  from first principles. One possible technique consists
in the direct measurement of the matrix element (\ref{eq:matrix}).
An alternative is to use the anomaly equation, which after taking the
divergence of both sides of Eq.(\ref{eq:matrix}), 
\begin{eqnarray}
 \langle  {\vec p'}, s' | Q | {\vec p},s \rangle &=& {m_N \over N_f}
\label{eq:array1}
 A(k^2) \overline{u}({\vec p'}, s') i \gamma^5 u({\vec p}, s) \\
 A(k^2) &=& G_1(k^2) + {k^2 \over m_N} G_2 (k^2) .
\label{eq:array2}
\end{eqnarray}
As $k \longrightarrow 0$ Eq.(\ref{eq:array2}) determines $G_1(0)$,
unless $G_2(k^2)$ has a pole at $k^2=0$ and this is the case in the
quenched approximation but not in full QCD. Eq. (\ref{eq:array1})
gives thus $\Delta\Sigma$ in terms of the matrix element 
$\langle  {\vec p'}, s' | Q | {\vec p},s \rangle$, which can
be measured on the lattice. In principle the lattice operator $Q_L$
would mix with $\partial_\mu J^5_\mu(x)$ and $\overline{\psi} \gamma^5
\psi$, but this mixing, as well as the small anomalous dimension of
$Q$ can be neglected~\cite{a4}. 
Our preliminary value is $\Delta\Sigma=0.04(4)$.
Here again the error could be larger and in any case the value
is preliminary, due to the bad sampling of topology in our ensemble
of configurations.

\section{Conclusions}
Measurement of the topological susceptibility $\chi$ on the lattice
is fully under control. For quenched $SU(3)$ the value is in good
agreement with the prediction of~\cite{witten,veneziano}. 
$\chi$ drops to zero at
the deconfining transition. Preliminary determinations of $\chi'$
in full QCD agree with sum rules. The spin content of the proton
is at hand. The practical problem is the thermalization of topology
on the lattice. Our huge sample of configurations is not thermalized
with respect to it. This creates in principle a problem for the 
lattice determination of any quantity: a priori, indeed, it is not
known how it could depend on the topological sector and therefore
if the ensemble is biased with respect to topology, this could 
affect the result in an impredictable way. Solutions of this
numerical problem are currently under study.

% ---- Bibliography ----
%

\end{document}